\documentclass[%
 reprint,
 amsmath,
 amssymb,
 aip,
 jcp
]{revtex4-1}

\usepackage{graphicx}
\usepackage{dcolumn}
\usepackage{bm}
\usepackage{hyperref}
\usepackage{xcolor}
\usepackage{tikz}
\usepackage{ulem}
\newcolumntype{?}{!{\vrule width 2pt}}

\usetikzlibrary{calc,decorations.markings}

\begin{document}

\preprint{APS/123-QED}

\title{Asymmetric and Long Range Interactions in Shaken Granular Media}

\author{Joan Codina}
\affiliation{Institute of Physics, Chinese Academy of Sciences, 100190 Beijing, China.}%
\affiliation{Wenzhou Institute, University of Chinese Academy of Sciences, 325001 Wenzhou, China.}%
\affiliation{%
	Departament de F\'isica de la Mat\`eria Condensada, Universitat de Barcelona, 
	C. Mart\'i i Franqu\`es 1, Barcelona 08028, Spain.}%

\author{Ignacio Pagonabarraga}%
 \email{ipagonabarraga@ub.edu}

\affiliation{%
	Departament de F\'isica de la Mat\`eria Condensada, Universitat de Barcelona, Mart\'i i
Franqu\`es 1, 08028 Barcelona, Spain.}%
\affiliation{%
Universitat de Barcelona Institute of Complex Systems (UBICS), Universitat de Barcelona, Barcelona 08028,
Spain.}
 \affiliation{CECAM Centre Europ\'een de Calcul Atomique et Mol\'eculaire, \'Ecole Polytechnique F\'ed\'erale de Lausanne, Batochimie, Avenue Forel 2, 1015 Lausanne, Switzerland.}
\date{\today}

\begin{abstract}
We use a computational model to investigate the emergence of interaction forces between pairs of intruders in a horizontally vibrated granular fluid. The time evolution of a pair of particles shows a maximum of the likelihood to find the pair at contact in the direction of shaking. This relative interaction is further studied by fixing the intruders in the simulation box where we identify effective mechanical forces, and torques between particles and quantify an emergent long range attractive force as a function of the shaking relative angle, amplitude, and the packing density of grains. We determine the local density and kinetic energy profiles of granular particles along the axis of the dimer to find no gradients in the density fields and additive gradients in the kinetic energies.
\end{abstract}

\maketitle


\section{\label{sec:intro}Introduction}
Granular matter has attracted recent attention because of its intrisically different nature {as} compared to equilibrium  solid, liquid and gas states~\cite{jaeger1996}; due to its ubiquity in the industry in the form of plastic beads, sand, as well as,  edible grains, \textit{i.e.}, coffee, wheat;
jamming and clogging processes, and stress distribution which lead to the formation of stable structures. {Jamming in granular matter} has been widely studied in the past~\cite{Liu1998,Cates1998}.

Shaken granular matter are actuated systems out-of-equilibrium due to a steady flux of energy from the container to the grains that is finally dissipated through contact interactions.
Striking phenomena occur in mixtures of particles where shaking leads to species segregation ranging from clusters to stripes~\cite{Ottino2000,Kudrolli2004,Aranson2006}.
The best known manifestation of this granular separation is the so called Brazil nut effect~\cite{Godoy2008,Ciamarra2006a,Sanders2004}.
For horizontally driven matter, gravity is no longer a relevant parameter, and a mixture can phase separate into stripes orthogonal to the shaking direction~\cite{Mullin2000,Reis2002,Reis2006,Ciamarra2007}, or even form clusters for swirling shakings~\cite{Aumaitre2001}.
Understanding the ability of granular mixtures to demix is of direct industrial relevance and can help to explain diverse phenomena, such as stratification in terrestrial environments or in asteroid or planetoid formation~\cite{opsomer2014}.

As a paradigmatic example of the qualitative change that granular systems experience under shaking we have the case of granular flow though a bottleneck -- like the flow of sand in an hourglass. Under the sole effect of a constant force, such as gravity, the granular flow can be spontaneously interrupted~\cite{Masuda2014,Lozano2012} if the exit is not wide enough. On the contrary, for granular matter under shaking the nature of the flux interruption dramatically changes and the system, once clogged, may spontaneously unclog~\cite{zuriguel2014}.  

In the non-equilibrium state induced by shaking, granular matter has been reported to fluidize for vertical and horizontal~\cite{Saluena1999,Saluena2000,Ristow1997} forcing. Once granular matter fluidizes, large density fluctuations have been reported and such a situation leads to the opening of Casimir-like scenarios~\cite{kardar1999} for the effective  granular interactions~\cite{cattuto2006}.

Band segregation for a binary mixture of horizontally agitated granular matter has been analyzed for attractive and anisotropic pairwise interactions between inclusions~\cite{Ciamarra2006b}.
However, the details of the interactions were not directly addressed.
Position tracking experiments of metallic spheres immersed in a dense poppy-seed granular monolayer have revealed not only an attraction between pair but an aligning interaction relative to the shaking direction~\cite{Lozano2015}. Following this experimental evidence we present this model that captures and quantifies the pair interaction between intrusive particles.


This paper is structured as follows. In Section~\ref{sec:model} we present a model for both grains, and inclusions under horizontal vibration.
In Section~\ref{sec:2d}, we study the free motion of an intruder pair pair in a shaken granular bed and identify the probability to locate particles as a function of their separation.
In Section~\ref{sec:fixed}, we fix the inclusion pair in order to quantify the emergent effective radial and tangential forces; the mechanical formation energy of the dimers is then compared to the pair distance distribution previously obtained in Section~\ref{sec:fixed}; and finally the effect of the pair in the kinetic energy of the granular bed.

\section{\label{sec:model}Model}
Recent experiments have studied the dynamics and structure of a pair of phosphor-bronze spheres immersed in a horizontally shaken monolayer of poppy seeds.
Poppy seeds are kidney shaped granular particles with typical diameters ranging $0.5~\text{mm}$ to $1~\text{mm}$ with material density $\rho=0.2~\text{g cm}^{-3}$.
Poppy seeds have a wide contact area with the surface of the tray, so their friction is large, and periodically displace with the vibrated tray.
Bronze spheres, on the contrary, are smooth, and uniform in size with a diameter of $1.5~\text{mm}$, and material density~\cite{Reis2004} $\rho=8.8~\text{g cm}^{-3}$.

\begin{figure}[h]
\includegraphics[width=1\columnwidth]{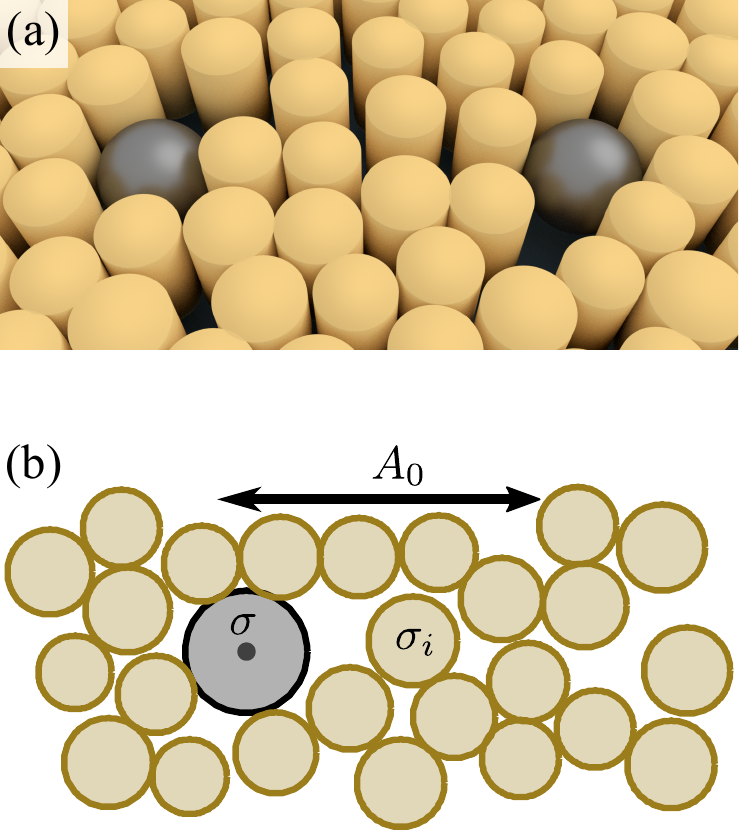}
\caption{\label{fig:sketch} Representation of the granular system. (a) Granular particles have a large contact area with the shaking surface and are modeled as cylinders whereas rotating inclusions are modeled as spheres. (b) Projection of the shaken granular system in the shaking plane.}
\end{figure}

Spheres easily rotate on the vibrated flat surface of the system~\cite{Reis2004}, whereas grains have a tendency to follow the moving tray.
Inspired by this situation, and as sketched in Fig.~\ref{fig:sketch}, we consider a simplified two dimensional model for the dynamics of grains and inclusions which captures the essential features of the forced grain monolayer, and develop an integration scheme to computationally solve the equations of motion of the granular system.

\subsection{\label{sub:eqs1}Equations of motion}
We treat grains as $N$ disks at positions $\bm x$ of diameter $\sigma$, drawn from a uniform distribution with average $\left\langle \sigma\right\rangle=\sigma_g$ and $10\%$ dispersion to account for experimental polydispersivity~\cite{Lozano2015}, in a 2d periodic box of side $L$.
We define an averaged packing fraction of the monolayer as $\phi=N\pi\langle \sigma^2 \rangle/(4L^2)$. Each grain with a mass that depends on their diameter as $m=m_0\left(\sigma/\sigma_g\right)^3$, where $m_0$ is the mass of a grain of size $\sigma_g$. Grains are periodically driven with the tray velocity $\bm v_s(t)$.
Bronze spheres, inclusions, are modeled as disks of diameter $\sigma=3 \sigma_g/2$, and mass $M=50\ m_0$ at positions $\bm X$. The solid rotation of inclusions on the moving plate is compatible with the description of the position vector $\bm X$ relative to the laboratory frame, \textit{i.e.}, we assume that inclusions are not being displaced by the tray motion $\bm v_s(t)$. For each particle, either $\bm x$ or $\bm X$ we present its corresponding equation of motion,
\begin{equation}
m_i\frac{d^2 {\bm x}_i}{dt^2} = - \gamma_{s,i} \left( \frac{d {\bm x}_i}{dt}-{\bm v}_s(t)\right) + \bm F_i^c + \bm F_i^d + \bm F_i^r~,
\label{eq:newton_grans}
\end{equation}
\begin{equation}
M\frac{d^2 {\bm X}_i}{dt^2} = - \gamma_s  \frac{d \bm X_i}{dt} + \bm F_i^c + \bm F_i^d + \bm F_i^r~,
\label{eq:newton_inclusions}
\end{equation}
where we introduce an oscillatory velocity of the tray, $\bm v_s(t) = A_0\omega \sin\omega t \hat{\bm e}_x$, characterized by its displacement amplitude, $A_0$, and frequency $\omega$.
The motion of the grains relative to the tray introduces a friction force whose magnitude $\gamma_s\left(d\bm x_i / dt-{\bm v}_s\right)$ depends on the velocity of the grain relative to the tray, and a dissipation constant $\gamma_s$ proportional to the contact area of the disk $\sigma^2$.

The excluded volume interaction among  particles, as well as the  energy dissipated in particle collisions, is described through a linear spring-dashpot model for the interactions between particles.
The total conservative and dissipative forces on particle $i$  are pairwise additive, ${\bm F}_i^{(c/d)}=\sum_j{\bm F}_{ij}^{(c/d)}$ that activate when grains overlap~\cite{Hidalgo2009,Saluena2000,poschel2005}.
The degree of compression between two disks $i$ and $j$ is quantified by  $\xi_{ij}=\left|{\bm x}_{ij}\right|-\left(\sigma_i+\sigma_j\right)/2$. Accordingly, the elastic pairwise conservative force reads
\begin{equation}
\bm F_{ij}^c = k \xi_{ij}\theta\left(\xi_{ij}\right)\hat{\bm n}_{ij}~,
\end{equation}
with $k$ an elastic constant (a material dependent parameter), and $\hat{\bm n}_{ij}$ the centre-to-centre unit vector.
$\bm F_{ij}^c$ does not vanish only when the two disks overlap, \textit{i.e.}, $\xi_{ij} > 0$ as accounted to by the Heavisde function, $\theta(\xi)$.

The dissipative pairwise interaction, responsible of the energy loss, reads
 \begin{equation}
\bm F_{ij}^d =\gamma \left({\bm v}_{ij}\cdot {\bm n}_{ij}\right)\theta\left(\xi_{ij}\right) {\bm n}_{ij}~,
\end{equation}
with $\gamma$  a dissipation constant,  $\bm v_{ij} = \bm v_i-\bm v_i$, and its direction opposes the direction of the relative velocity between pairs.
This dissipation mechanism between pairs may lead to configurations of small overlaps with the elastic and dissipative forces canceling each other. In these situations pairs of particles tend to remain close to each other. To avoid such computational artifacts, as proposed in~\cite{poschel2005}, we use the total force of interaction between pairs to be always either repulsive or zero $|\bm F_{ij}| = \max(0,|\bm F^c_{ij} + \bm F^d_{ij}|)$.
 
Additionally, we introduce random forces, $\bm F^r$, to account for irregularities and vertical collisions. The source of noise is Gaussian for each component with zero mean and variance $\left\langle \hat{\bm e}_\alpha\cdot \bm F^r_i(t)\hat{\bm e}_\beta\cdot \bm F^r_j(t')\right\rangle = 2\Lambda_\alpha \delta_{ij}\delta_{\alpha\beta} \delta\left(t-t'\right)$.
The $\Lambda_\alpha$ parameter accounts for the asymmetry in the noise strength in the parallel and perpendicular directions relative to the tray shaking.

In the following sections we will consider the tray  oscillation  period $\tau^{-1}= 2\pi\omega$, the  average grain size  $\sigma_g$, and its mass $m_0$ as time, length and mass units respectively. In terms of these magnitudes we define the elastic constant $k$, the dissipation constants $\gamma$, and $\gamma_s$, and $\Lambda_x$ with standard values present in the literature\cite{Saluena1999}.
The integration of the equations is performed by a stochastic integrator inspired on~\cite{allen1989} and detailed in Appendix~\ref{ap:sdg}.

In Appendix~\ref{ap:densitats} we characterize a bed of granular particles and analyze their kinetic energies. We observe that the packing fraction of the grain monolayer has an impact in the inclusion energy loss above $\phi=0.6$ whereas beyond $\phi\approx 0.75$ the dynamics of the grains slows down and the the displacement statistics of grains show the caging effects introduced by the formation of a quasi-crystalline structure of the grains.
In the rest of this paper we consider granular beds within the density range $\phi\in [0.6,~0.75]$.

\section{\label{sec:2d} Free moving dimer}
In order to analyze the emergent features induced by a forced granular system on  embedded intruders, we will analyze the dynamics of a pair of intruders. 
To this end, we consider the free evolution of a dimer of inclusion particles in a granular bed. In subsection~\ref{sub:probabilities} we define the relative coordinates that describe the pair motion.
In subsection~\ref{sub:landscapes} we extract the relative arrangement of the inclusion pair from its evolution as a function of the  granular shaking amplitudes and packing fractions.
Finally, in~\ref{sub:compression} we analyze the the averaged probability distribution of the pair relative positions to measure the pair separated a distance $d$ and the relative orientation for both touching and distant pairs.
\subsection{\label{sub:probabilities} An inclusion dimer}
In a periodically driven granular system the energy is introduced by the external forcing through both  $A_0$, and $\omega$.
On the one hand, the shaking amplitude $A_0$ introduces a length scale that may lead to structural system deformations of typical length $\sim 2A_0$.
On the other hand, the shaking frequency $\omega$ modifies the dissipation rate of the granular system.
In this paper we control the tray vibration by changing the shaking amplitude, $A_0$, and fixing its frequency.

A doublet of intruders of diameter $\sigma$, $\{\bm X_a,~\bm X_b\}$, defines a dimer whose center to center vector is ${\bm r} = \bm X_a - \bm X_b$.
The position vector ${\bm r} =(r_x,~r_y)$ is equally described by its modulus $r$ and angle $\alpha$ relative to the shaking direction, $\bm r = r\left(\cos\alpha,~\sin\alpha\right)$.
We define a ``parallel'' (``perpendicular'') configuration of the dimer for a configuration $\alpha\approx 0$ ($\pi/2$).
For simplicity we also use the distance between the surfaces of inclusions $d = r - \sigma$ to define the dimer distance.

\subsection{\label{sub:landscapes} Probability Landscapes}
In order to understand the dynamics of a dimer of intruders  in a shaken granular bed we have prepared a large collection of initial conditions with $d\in\left[4,7\right]$ and $\alpha\in\left[0,2\pi\right]$ to sample between $10^6$ ($\phi=0.6$) and $6\cdot 10^6$ ($\phi=0.75$) cycles for systems of box size $L=32\sigma_g$ at shaking amplitudes $A_0=0.75\sigma_g$ and $A_0=1.5\sigma_g$.

We define $\mathcal{P}(r_x,~r_y)$ as the probability to measure the dimer in a configuration with $r_x\in \left[r_x, r_x+\Delta x\right]$ and $r_y\in \left[r_y,~r_y+\Delta y\right]$, where $\Delta x=\Delta y=0.05\sigma_g$, that results from simulations sampling at each shaking cycle.
From this probability, we can extract $-\ln \mathcal{P}\left(r_x,r_y\right)$, which identifies the most probable configurations of the dimer freely evolving in the granular bed.
If the system were in equilibrium, the logarithm of the probability $-\ln \mathcal{P}$ would be proportional to the energy landscape $\mathcal{U}(r_x,r_y)$ of the system. 
We leave the comparison between $\mathcal{P}$ and the dimer mechanical formation energy, $U$, for section~\ref{sub:energy}.
\begin{figure}[h]
\includegraphics[scale=1]{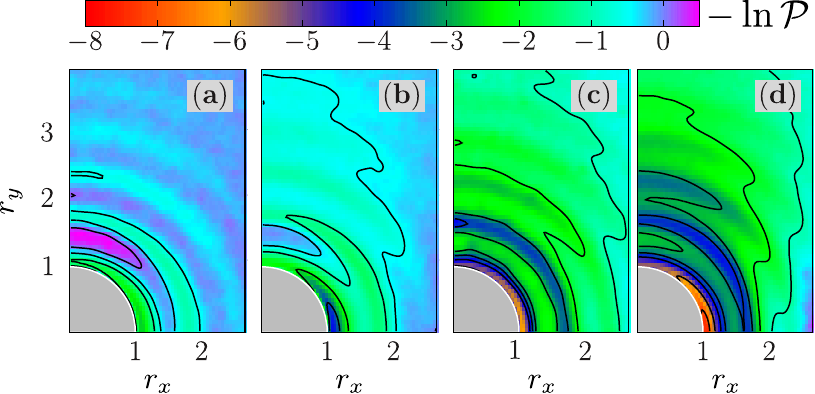}
\caption{\label{fig:2dmap} Landscape of dimer configurations. Heat map with solid contour lines of $-\ln\mathcal{P}$ as a function of the relative configuration of the dimer, $(r_x,~r_y)$, for different grain packing fraction, $\phi$, and forcing strength, $A_0$.
The gray shaded regions correspond to configurations of $r_x^2+r_y^2< \sigma$, and thus cannot be reached by the inclusion.
The x-component  is aligned with the sinusoidal forcing imposed on the tray.
From left to right: (a)-(b) $\phi=0.6$ and (a) $A_0=0.75\sigma$, and (b) $A_0=1.5\sigma$;  (c)-(d) $\phi=0.75$ and (c) $A_0=0.75\sigma_g$, and (d) $A_0=1.5\sigma_g$.} 
\end{figure}
We associate minima of the magnitude $-\ln P$ to more probable configurations (effective attractions), and maxima to more unlikely configurations (effective repulsions).
Probability landscapes $-\ln \mathcal{P}(r_x,r_y)$, see Fig.~\ref{fig:2dmap}, identify $r\approx \sigma$ as the most probable pair distance for all studied granular beds.
Additionally, all four baths show a high degree of anisotropy at contact, $r\approx\sigma$, where $-\ln \mathcal{P}(\sigma, \alpha)$ is always smaller for $\alpha\approx 0$ than for $\alpha\approx \pi/2$. This clear asymmetry shows a strong tendency of dimer to align in the shaking direction.

The structure of the granular bed is clearly manifested in the probability landscape as a sequence of concentric oscillating rings.
At long distances from contact, contour lines manifest the anisotropy of the interaction.
Moreover, the probability landscape decays faster along the shaking direction than perpendicular to it.

\begin{figure}[h]
\includegraphics[scale=1]{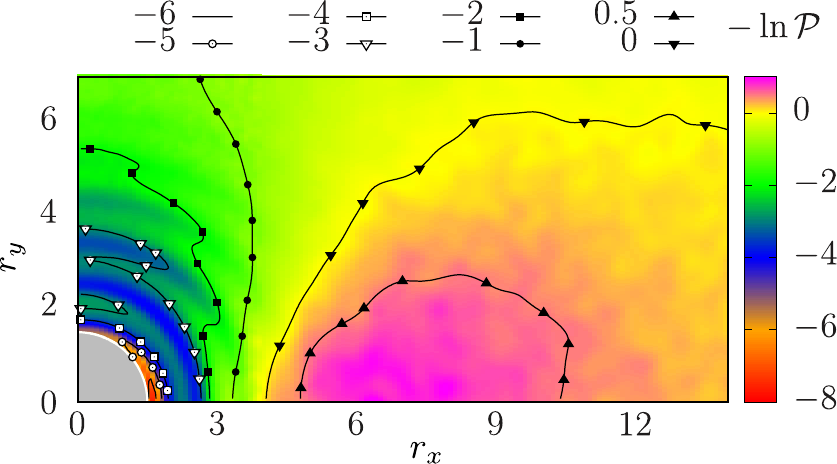}
\caption{\label{fig:anisotropy} Anisotropy in the dimer configurations. Heat map with solid contour lines of the landscape, $-\ln\mathcal{P}$ for a system at $\phi=0.75$, and $A_0=1.5\sigma_g$, an extension in the $\bm r$ domain of Fig.~\ref{fig:2dmap} to show the long distance behavior. To fully comprehend the behavior of $-\ln\mathcal{P}$ we introduce different symbols in the contour lines, as detailed in the upper key. We observe an exclusion zone, in magenta, where the probability to find the dimer is below the value of $\mathcal{P}$ at infinity.}
\end{figure}

Fig.~\ref{fig:anisotropy} further details the probability landscape for  $A_0=1.5\sigma_g$, and packing fraction $\phi=0.75$.
The granular structure  is rapidly lost in the parallel direction whereas it shows stronger persistence for configurations where the dimer is oriented perpendicular to the tray shaking direction.
At long distances, we observe that perpendicular configurations are favored and an island of repulsion emerges centered at $r=8\sigma_g$, and $\alpha=0$.

\subsection{\label{sub:compression} Marginal probability densities}
We can provide a more compact description of the distance and angular dependence that characterize the dimer interaction through the corresponding marginal probability densities.
These quantities are also more attractive experimentally because they require less statistics.
Specifically, we quantify the distance behavior of $\mathcal{P}$ through   $P(r)$, the averaged angular distribution, obtained integrating  $\mathcal{P}(r,\alpha)$ over the angular configurations of the dimer.

\begin{figure}[h]
\centering
\includegraphics[scale=1]{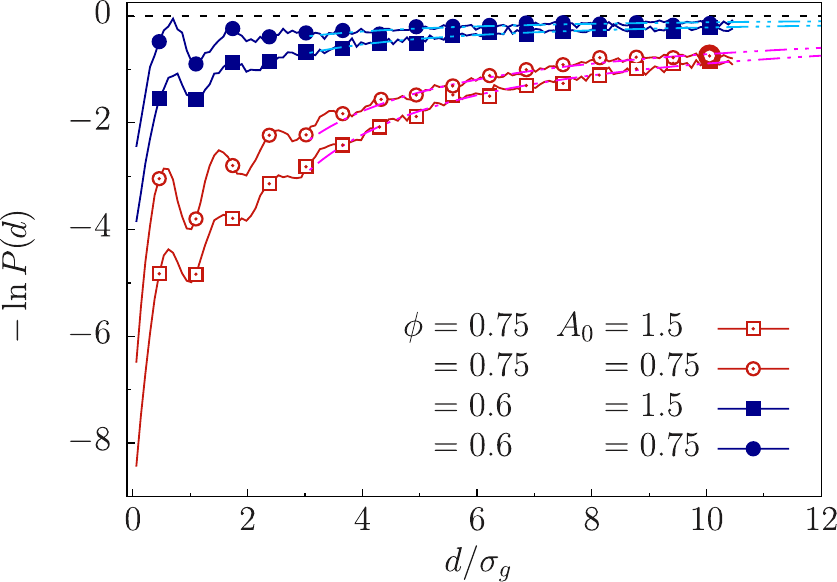}
\caption{\label{fig:radial_potencial} Angular averaged probability density. Plot for different values of the packing density $\phi$, and shaking amplitude $A_0$.
As a guide to the eye, we plot the fitted  $-\ln P(d) = a/d$ in cyan, and magenta.}
\end{figure}

Fig.~\ref{fig:radial_potencial} displays the angular averaged probability density, $P(d)$,  as a function of $d=r-\sigma$, so that $d=0$ corresponds to the contact position $r=\sigma$.
We identify three different features in $P(d)$.
First, at contact ($d=0$), $-\ln P(d)$ reaches its lowest value; hence it is the preferred and most probable configuration.
Second, the oscillations of $-\ln P(d)$ at short distances clearly show the structure of the bed since the periodicity corresponds to the granular size.
The surroundings of minima are locally stable and correspond to an integer number of grains in the internal region of the inclusion dimer. Third, the oscillations are superimposed on a monotonous decay towards zero.
The decay is stronger the more compact the system is -- the signal is stronger at $\phi=0.75$.
Within the available range of distances, we have tested that the decay in $-\ln P(d)$ is compatible with an algebraic tail $-\ln \big(P(d)\big/P(\infty)\big) \sim d^{-1}$, shown in dashed lines in Fig.~\ref{fig:radial_potencial}.

To quantify anisotropy we compute the {conditional probability densities} $P_<(r^\star;\alpha)$, and $P_>(r^\star;\alpha)$.
Obtained as integrals of $\mathcal{P}(r,\alpha)$ over $r$ in the ranges $r\in\left(\sigma,\sigma+r^\star\right)$, and $r\in\left(\sigma + r^\star,\infty\right)$ for $P_<$ and $P_>$ respectively.
$P_<(r^\star;\alpha)$ quantifies the anisotropy of the configurations of the dimer at short distances whereas $P_>(r^\star;\alpha)$ quantifies long distance anisotropy.


\begin{figure}[h]
\begin{center}
\includegraphics[scale=1]{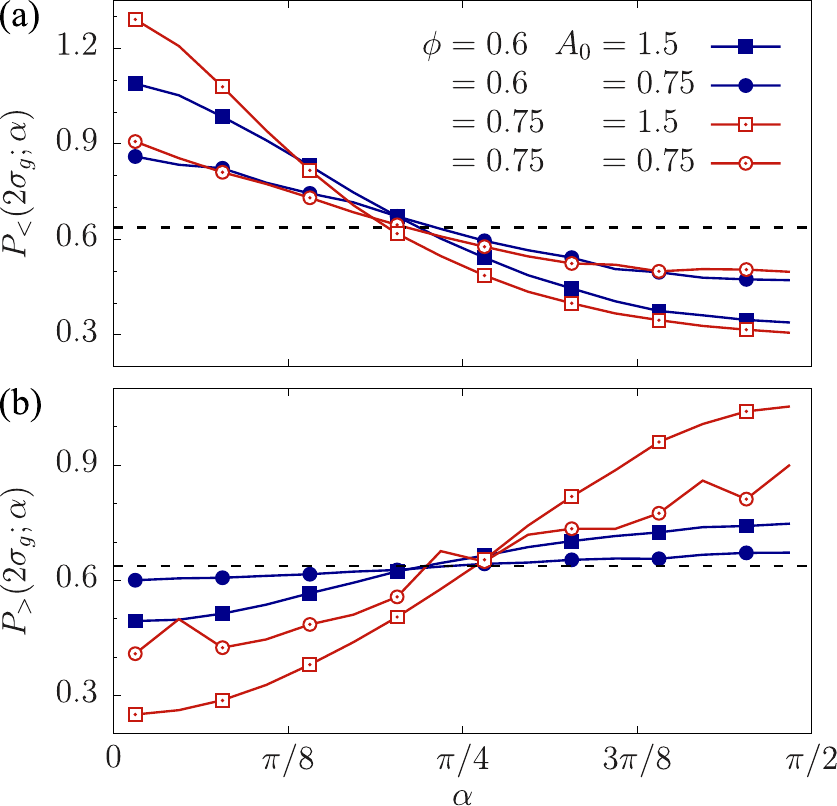}
\end{center}
\caption{\label{fig:dist_angles} Angular dependence of the conditional probability densities as a function of $\phi$, and $A_0$. (a) for distances $d< 2\sigma_g$.
(b) for distances $d>2\sigma_g$. Black dashed lines display the value of an isotropic distribution, $2/\pi$.}
\end{figure}

The angular distributions displayed in Fig.~\ref{fig:dist_angles} have been obtained for $r^\star = 2\sigma_g$.
At short distances, $P_<(2\sigma;\alpha)$, we observe a preference for parallel configurations.
The anisotropy in the configurations has a major dependence on the shaking amplitude, while it shows a mild dependence on packing density.
At larger distances, $d>2\sigma_g$, on the contrary, the dimer tends to orient perpendicular to the shaking direction.
In this case the degree of anisotropy is relatively weak except for large densities and shaking amplitudes, $\phi=0.75$ and $A_0=1.5\sigma$, where the anisotropy increases revealing the instability island clearly appreciable in the 2d heat map in Fig.~\ref{fig:anisotropy}.

If the system were to be in equilibrium, the interpretation of the probabilities would be straightforward.
The minimum, and slow decay of the radial probability $-\ln P(d)$ would translate into an energy minimum at $d=0$, and an emergent attractive long range interaction $U\sim d^{-1}$.
The results in the anisotropy shown in Fig.~\ref{fig:dist_angles}(a) would imply the emergence of an aligning torque at short distances; a torque that would align the dimer towards the shaking direction. However, the system is not in equilibrium, and thus we cannot relate the probability of the dimer configurations to the interaction energies in the system.
In the following section we compute mechanically the effective interaction forces and formation energies of a dimer following similar procedures to the ones used by \cite{harder2014,Ni2015,zaeifi2017}.

\section{\label{sec:fixed} Fixed inclusions}
In this section we quantify the mechanical interactions between inclusions that arise from the granular shaking.
Fist in~\ref{sub:interaction} we fix the inclusions at relative coordinates $(r,~\alpha)$ and measure their relative mean radial and tangential interactions.
Second in~\ref{sub:angle} we propose a model that captures the long distance behavior of the radial force as a simple function of the shaking amplitude $A_0$, the relative angle $\alpha$, and the inclusions separation distance $d$.
Third in~\ref{sub:energy} we define the mechanical formation energy of the dimer and compare it to the configurations' probabilities for the freely evolving dimers measured in Section~\ref{sub:compression}
Finally in~\ref{sub:profiles} we compute local profile measures of the granular bed along the axis of the dimer to provide further insight in  the mechanism underlying the effective interactions between inclusions.

\subsection{\label{sub:interaction} Radial and tangential forces}
We fix the pair of inclusions at positions $\bm X_a$ and $\bm X_b$ but we let the velocity of the inclusions evolve in time since particle collisions depend on the relative velocity between interacting particles.
The forces given by the granular bed on the inclusions are $\bm F_a$ and $\bm F_b$ and define the effective interaction between inclusions.
Computing the relative force $\bm F_b(\bm X_b) - \bm F_a(\bm X_a)$ we extract the relevant information of the relative force acting on the dimer.
Finally we project along the radial and tangential directions,
\begin{equation}
F_r(d) = \left \langle \frac{\bm F_b(\bm X_b) - \bm F_a(\bm X_a)}{2}\cdot \hat{\bm r} \right\rangle
\label{eq:force_r},
\end{equation}
\begin{equation}
F_t(d) = \Big\langle \big(\bm F_b(\bm X_b) - \bm F_a(\bm X_a)\big)\cdot \hat{\bm t}  \Big\rangle 
\label{eq:force_t},
\end{equation}
where unit vectors $\hat{\bm r}$ and $\hat{\bm t}$ are an orthonormal basis of $\mathbb{R}^2$ with $\hat{\bm r}=\left(\bm X_b-\bm X_a\right)/\left|\bm X_b-\bm X_a\right|$ and, after rotating $\pi/2$ we obtain $\hat{\bm t}=\mathcal{R}_{\pi/2}\hat{\bm r}$.
The brackets $\left\langle \cdot\right\rangle$ denote the average of the magnitude $(\cdot)$ over cycles and independent realizations of the system.

Inclusions have been placed at configurations with surface to surface distance the range $d\in \left[0, 8\right]$ and angles relative to shaking $\alpha \in \left[0,\pi/2\right]$.
We average on various sets with fixed parameters $\alpha, d, A_0, \phi$ to extract the mean values $F_r$, and $F_t$.

Within this definition of $\hat{\bm r}$, and $\hat{\bm t}$, the projection $F_r$ defines the behavior of the interaction in the radial direction with attractive effective forces captured by negative values of the relative force, $F_r<0$, and repulsive otherwise.
The transverse effective force, $F_t$, indicates the emergence of a neat torque acting on the dimer with clock-wise torques identified by $F_t<0$, and counterclock-wise otherwise. For our dimer configurations in the first quadrant of the plane, clock-wise tangential forces align the dimer in the shaking direction.

\begin{figure}[b]
	\includegraphics[scale=1]{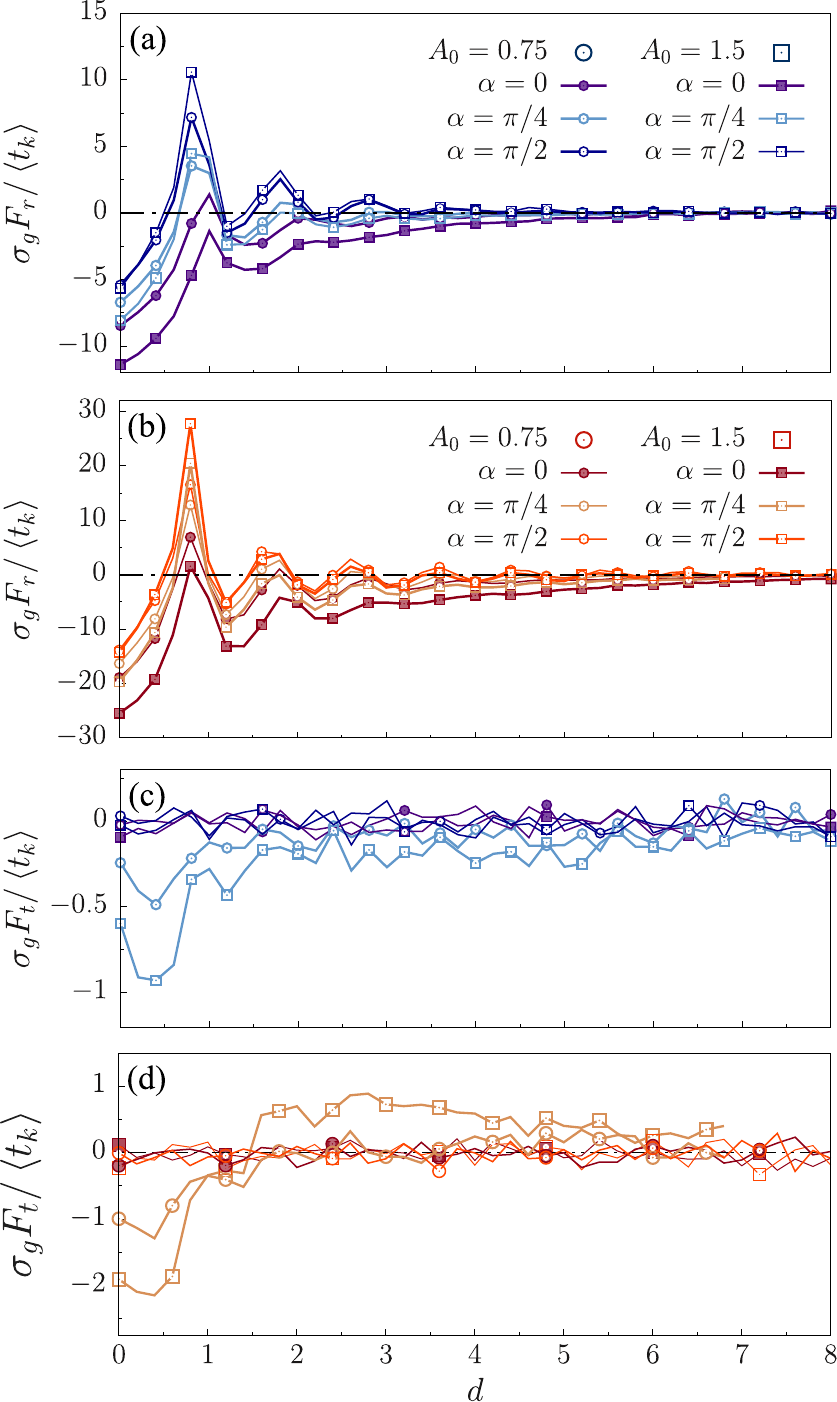}
	\caption{\label{fig:forces} Effective radial, and tangential forces  $F_r$ in (a, b),  $F_t$ in (c, d), respectively. Plots (a, c) for density $\phi=0.6$ and (b, d) for density $\phi=0.75$. Different amplitudes are labeled by squares, $A_0=1.5\sigma_g$, and circles $A_0=0.75\sigma_g$. Different shades of colors identify the dimer angle relative to the shaking direction.}
\end{figure}

The relative radial force $F_r$ in Fig.~\ref{fig:forces} reflects that the effective interaction for a pair of inclusions at contact is attractive.
We observe that the magnitude of the force at contact also depends on the orientation of the dimer relative to the shaking direction.
The interaction is stronger the more aligned --to the shaking-- the dimer is.
As distance increases, the force reaches a local maximum before $d=\sigma_g$ and then $F_r$ oscillates with an amplitude that reflects the structure of the grain suspension.

The magnitude and sign of the radial force is sensitive to the orientation of the dimer with respect to the tray shaking direction, $\alpha$.
For perpendicular configurations the decaying oscillations in the force reflect the structure of the granular bed for long distances as it is clear from the persistence of the force oscillations.
As the angle is reduced, $\alpha<\pi/2$, the oscillation structure in the radial force is gradually destroyed at long distances while a neat attractive force remains for several inclusion diameters.
Overall, shaking destroys the structure of the granular bed and induces an effective attraction that increases as the dimer aligns relative to the shaking.

In addition to the radial interaction, a torque captured as a relative tangential force $F_t$ appears for configurations with neither $\alpha\neq0$, nor $\alpha\neq \pi/2$.
We have observed a maximum value in the tangential force strength for $\alpha=\pi/4$, which we select to display in Fig.~\ref{fig:forces}.
This non-central force is present in the system for short separation distances, $d\lesssim 2\sigma_g$ and orients the  pair along the shaking direction.
At $\phi=0.75$, and $A_0=1.5$ a finite dealigning torque clearly emerges at long distances $d\gtrsim 2\sigma_g$.
The nature of the tangential force shown in Fig.~\ref{fig:forces} is compatible with the anisotropy found in the probability $\mathcal{P}(r_x,r_y)$ for moving inclusions reported in Figs.~\ref{fig:2dmap} and~\ref{fig:dist_angles} in the previous section~\ref{sec:2d}.

\subsection{\label{sub:angle} Long distance behavior of the radial force}
\begin{figure}[h]
	\includegraphics[width=\columnwidth]{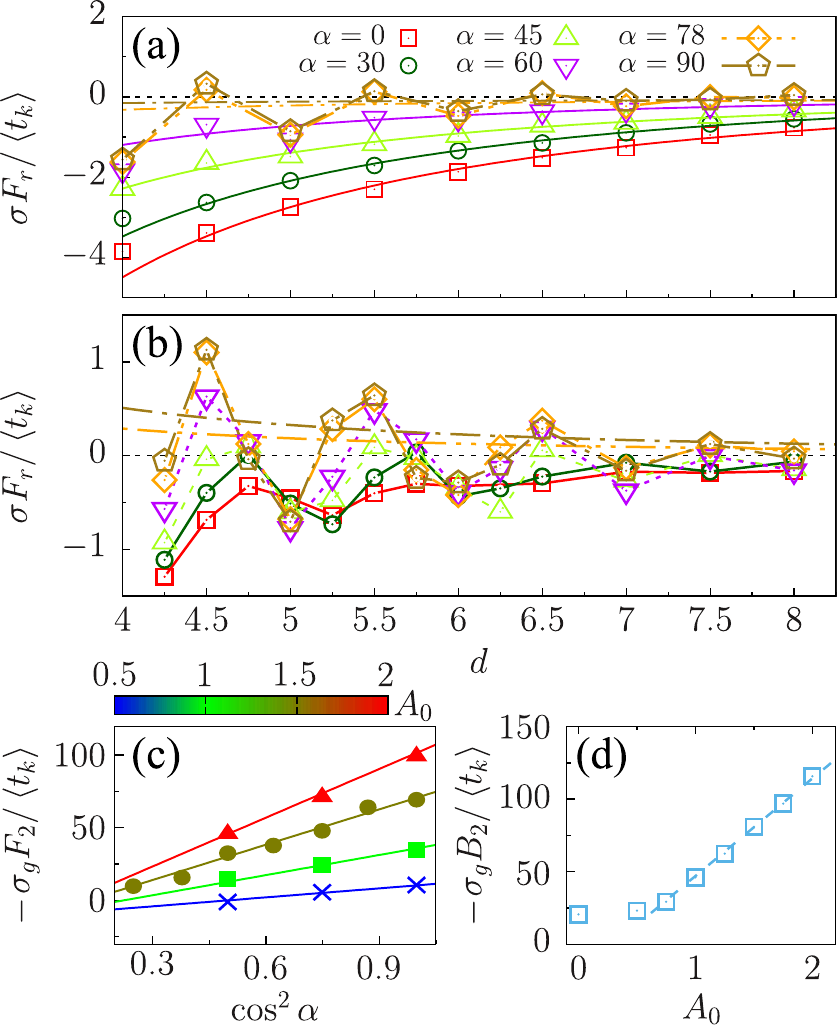}
	\caption{\label{fig:long_range} Long range interaction $F_r$ as a function of interparticle distance, angle, and shaking amplitude. (a) at amplitude $A_0=1.5\sigma_g$ the external forcing is strong enough to clearly destroy the internal structure of the gains at $\alpha<60^\text{o}$. (b) at $A_0=0.5\sigma_g$ the internal structure persists for angles $\alpha \gtrsim 30^\text{o}$. (c) Strength of the long range force $F_2$ as a function of $\cos^2\alpha$ for different shaking amplitudes. (d) Value of the prefactor of $\cos^2\alpha (\sigma_g/d)^{2}$ as a function of $A_0/\sigma_g$.}
\end{figure}

Fig.~\ref{fig:radial_potencial} shows that the effective force between inclusions generically displays an attractive, long-range interaction in length scales larger than $10\sigma_g$.
Even though we are not able to analyze the functional dependence of this asymptotic decay over several decades, the observed decay is consistent with an algebraic dependence.
In granular media, long range forces  arising from their non-equilibrium fluctuations have already been described in vibrated systems~\cite{cattuto2006}.
For the weak decaying signal of the radial force measure for configurations aligned close to the perpendicular direction an algebraic fitting curve of the form $F_2 (\sigma_g/d)^{2}$ has proven to be numerically more robust than an exponential decay which is largely interfered by the strong oscillations, see Fig.~\ref{fig:long_range}(a).
For an algebraic fitting, the strength of the long range interaction is given by the force $F_2$ and presented in Fig.~\ref{fig:long_range}.

To capture the full dependence on $\alpha$, and $A_0$ on the long range interaction  amplitude, $F_2$, we have systematically swept the space of parameters $\alpha$, and $A_0$.
The structure of the bed of grains is appreciable in both Fig.~\ref{fig:long_range} (a) and (b); at perpendicular configurations the emergent long range interaction vanishes, $F_2 \approx 0$.
At low values of the shaking amplitude, the grain bed structure persists at lower angles --denoted in dashed lines.
As the angle of the dimer decreases the effect of the external forcing increases and the local granular structure is lost.
In this regime a clear  interaction between particles develops, and increases as the dimer aligns towards the forcing direction.\

To quantify the dependence on the orientation $\alpha$ we present $F_2$ as a function of $\cos^2\alpha$, $F_2=B_2\cos^2\alpha+B_0$, the lowest power of $\cos\alpha$ allowed by the dimer and shaking symmetry.
Results in Fig.~\ref{fig:long_range}(c) confirm a quadratic dependence of the cosine on $F_2$. The analysis of $B_2(A_0)$ in Fig.~\ref{fig:long_range}(d) a $B_2$ that remains essentially constant for shaking amplitudes below a value $A_0^c$ followed by a linear increase for $A_0>A_0^c$,

Overall, we model the radial dimer interaction by means of the simple function,
\begin{equation}
F_r=F(\phi)\left[\frac{\theta(A_0-A_0^c(\phi))}{\sigma_g}+B_0(\phi)\right]\left(\frac{\sigma_g}{r}\right)^2\cos^2\alpha.
\label{eq:independence}
\end{equation}

For a system at packing fraction $\phi=0.75$ we obtain $A_0^c(\phi=0.75)=\left(0.6\pm 0.15\right)\sigma_g$ as the shaking amplitude above which the interaction increases linearly with $A_0$, $B_0(\phi=0.75)=0.3$, and $F=-67 \left\langle t_k \right\rangle /\sigma_g$, where we have prescribed an \textit{a priori} dependence on the packing density on each of the parameters.

\subsection{\label{sub:energy} Mechanical formation energies}
We benefit from the previous section, and in particular from the radial force displayed in Fig.~\ref{fig:forces} to define the mechanical formation energy of a dimer aligned an angle $\alpha$ with respect to the imposed external driving, $\mathcal{U}_\alpha(d)$.
Specifically, we define the dimer formation energy as the total mechanical work to bring a pair of inclusions from infinity to a relative distance $d$ along a line at constant angle,
\begin{equation}
\mathcal{U}_\alpha(d) = \int_{\infty\rightarrow d} \bm F_{int} \cdot d\bm l  = \int_d^\infty F_r(r,\alpha) dr.
\end{equation} 

\begin{table}[t]
	\caption{\label{tau:compararison} Effective temperature obtained by minimizing $\chi^2$ combining measures of the mechanical formation energy and the for different densities, shaking amplitudes, and orientation angles. Values with missing $\alpha$ correspond to the angular averaged formation energies and probabilities.}
	\begin{ruledtabular}
		\begin{tabular}{ccc c   cccc}
			$A_0$  &  $\phi$ &  $\alpha$ & $k_BT_{\mbox{eff}}$ &
			$A_0$  &  $\phi$  &  $\alpha$ & $k_BT_{\mbox{eff}}$\\
			\colrule 
			$0.75$  & 0.6    & $0$      & 0.35 &
			$0.75$  & 0.75    & $0$     & 0.24 \\
			0.75  & 0.6    & $\pi/4$    & 0.64 &
			0.75  & 0.75    & $\pi/4$   & 0.42 \\
			0.75  & 0.6    & $\pi/2$    & 7.4 &
			0.75  & 0.75    & $\pi/2$   & 1.21 \\	
			0.75  & 0.6    & $-$        & 0.61 &
			0.75  & 0.75    & $-$       & 0.46 \\
			1.5  & 0.6    & $0$        	& 0.35 &
			1.5  & 0.75    & $0$        & 0.16 \\
			1.5  & 0.6    & $\pi/4$     & 0.46 &
			1.5  & 0.75    & $\pi/4$    & 0.26 \\
			1.5  & 0.6    & $\pi/2$     & $-0.47$ &
			1.5  & 0.75    & $\pi/2$    & $0.95$ \\
			1.5  & 0.6    & $-$       	& $0.39$ &
			1.5  & 0.75    & $-$        	& $0.21$ 
		\end{tabular}
	\end{ruledtabular}
\end{table}

In equilibrium the energy of formation is derived from the probability to measure the pair at a given configuration $(d,~\alpha)$ by the simple expression $\mathcal{U}=- k_BT \ln {\mathcal P}$.
This relation implies, in particular, that the energy of formation scales with the system temperature, $k_BT$.
Since the granular bed is subject to an energy flux, induced by the tray shaking, we cannot take the previous equality for granted.
We subsequently test the validity of this previous relation through an effective bath temperature\cite{Song2005}, $k_B T_{\mbox{eff}}$ that could connect the formation energy to the probability to measure a dimer configuration.
\begin{figure}[h]
	\includegraphics[scale=1]{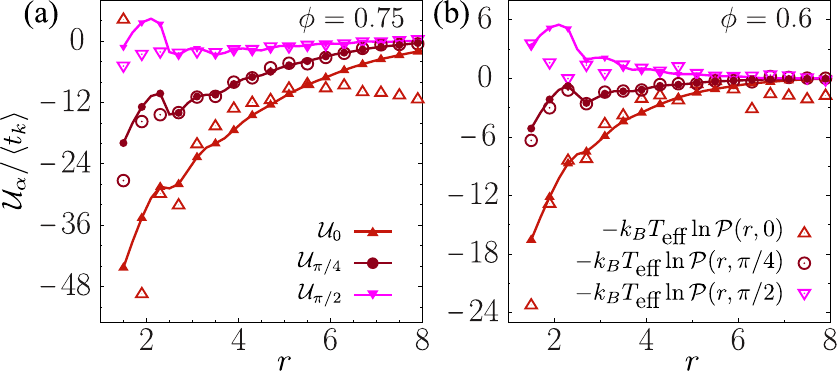}
	\caption{\label{fig:energy_angles} {Mechanical formation energy $\mathcal{U}_\alpha$ in solid lines and the correspondent effective potentials as obtained after minimizing $\chi^2$ for the probability densities $\mathcal{P}(r,\alpha)$ restricted at $\alpha$. We present shaking amplitudes $A_0=0.75\sigma_g$ (a), and $A_0=1.5\sigma_g$ (b) at $\phi=0.75$ with the best fits for the effective temperature $k_BT_{\mbox{eff}}$ as listed in Table~\ref{tau:compararison}}.}
\end{figure}
To this end, we fit the formation energy of the dimer to the probability to measure the pair of moving inclusions in that dimer configuration.
To extract an effective $k_BT_{\mbox{eff}}$ for each set of formation energies $\{\mathcal{U}_\alpha(d_i)\}$ and probabilities $\{\mathcal{P}(d_i,\alpha)\}$ defined at distances $\{d_i\}$ we minimize the chi squared function $\chi^2 = \sum_i \left[\mathcal{U}_\alpha(d_i) - \left( k_BT_{\mbox{eff}} \ln \mathcal{P}(d_i,\alpha) + U_0\right)\right]^2$, with $k_BT_{\mbox{eff}}$, and $U_0$ the fitting parameters.

Fig.~\ref{fig:energy_angles} compares the formation energy derived from the relative force and from the angular constrained probabilities with the best fit for $k_BT_{\mbox{eff}}$.
The corresponding temperatures depend in particular on $\alpha$, which indicates it is impossible to describe the dimer behavior as if it was immersed in an effective equilibrium bath, {see Table~\ref{tau:compararison}}.
Specifically, the obtained formation energies display qualitatively different dependence on the dimer distance when compared to $-k_BT_{\mbox{eff}}\ln P(d)$.
For perpendicular configurations the formation energy of the dimer presents a maximum with $\mathcal{U}>0$ that decays to zero. The minus logarithm of the probability, instead, presents an absolute minimum and then increases asymptomatically to zero. These qualitative differences do not allow to describe this system in terms of a $T_{\mbox{eff}}$.

\begin{figure}[h]
	\includegraphics[width=\columnwidth]{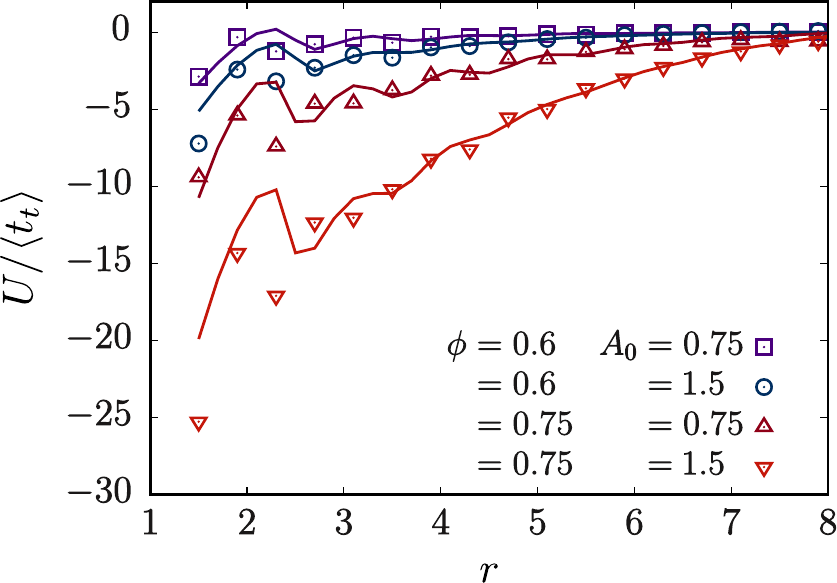}
	\caption{\label{fig:energy_fit} {Angular averaged formation energy $\left\langle U\right\rangle_\alpha$ in solid lines compared to the fitted values for $-k_B T_{\mbox{eff}} \ln P(d)$ in points. The values of the corresponding of $k_B T_{\mbox{eff}}$ are given in Table~\ref{tau:compararison} and labeled by $\alpha -$ }.}
\end{figure}

Alternatively, we can extract an effective temperature from the angular averaged probability, $P(d)$, see Fig.~\ref{fig:radial_potencial}.
To this end, we introduce the angular averaged mechanical formation energy, $\left\langle \mathcal{U}(d)\right\rangle_\alpha$, obtained integrating  $\left\langle F_r\right\rangle_\alpha \equiv 2/\pi \int_0^{\pi/2}d\alpha~ F_r(d,\alpha)$. Fig.~\ref{fig:energy_fit} shows that the formation energy obtained by the two routes shows semiquantitative agreement, and also, by construction, we can assign a $T_{\mbox{eff}}$ for each system. Hence, even if the comparison may be  misleading, it is possible to interpret the angular average quantities in terms of an effective equilibrium.

If the functional proposal holds, see Eq.~\ref{eq:independence}), the angular averaged  formation energy coincides with the  one obtained  from the general situation at $\alpha=\pi/4$, and that $-\ln P(d)\sim d^{-1}$ asymptotically. This functional dependence allows a fit between  $-\ln P(d)$, and $\left\langle \mathcal{U} \right\rangle_\alpha$ at long distances. 


\subsection{\label{sub:profiles} Measure profiles in the granular bed}
We analyze the impact that the inclusions have in the granular bed.
Fixing the inclusions allows us not only to measure the relative forces and torques between them, but also quantify the spatial dependence of the physical properties of the grains.
We focus on the relative variations of the granular properties along the axis that joins the two dimers.
To measure this profile, we consider a rectangular box of size $L\times\sigma_g$ that encloses the inclusions in the axis of the dimer.
We subsequently slice this box in cells of width $\Delta x=0.2\sigma$ and height $\Delta y=0.2\sigma$ and use them to build the corresponding histograms.

We compute the local density, $\phi(x',y')$, and kinetic energies $e_k(x',y')$, and $ t_k(x',y')$ at each cell with center at $(x',y')$ and average over oscillation cycles for different initial conditions of the sea of grains.
The result is then integrated in the vertical dimension to obtain the dependence on the distance from the surface of an inclusion $h$.
The pair of inclusions define an internal region which is identified as $-d/2<h<0$, and an external region $h>0$.
To gain statistics, we combine the results from the profiles centered on each inclusion.
For visualization purposes we plot the negative region from surface to surface ($-d<h<0$).
\begin{equation}
\delta e_k(h) = \frac{e_k(h) - \left\langle e_k\right\rangle }{\left\langle e_k\right\rangle},\quad \delta t_k(h) = \frac{t_k(h) - \left\langle t_k\right\rangle }{\left\langle t_k\right\rangle}
\label{eq:de}
\end{equation}

\begin{figure}[h!]
	\includegraphics[width=.95\columnwidth]{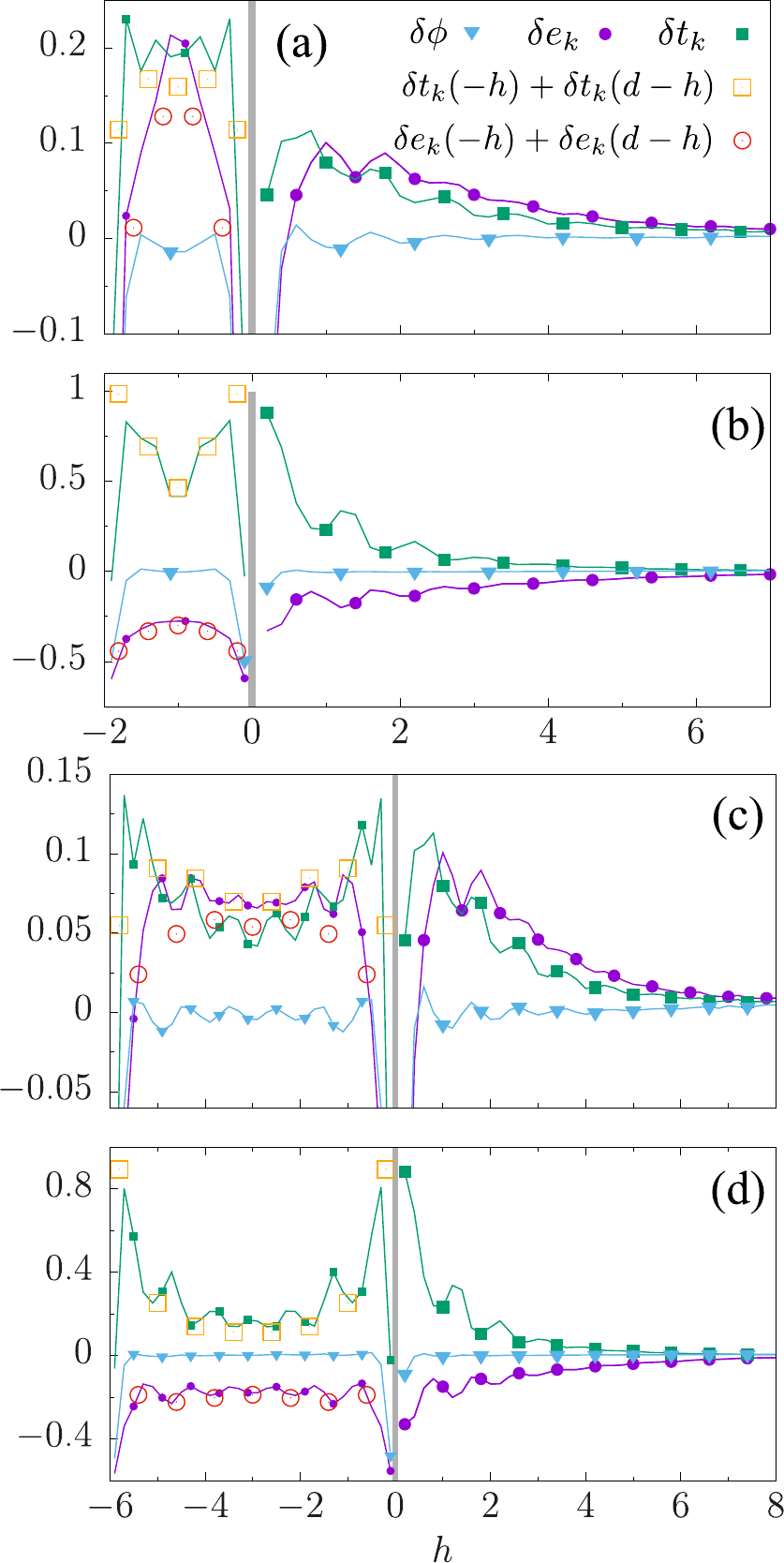}
	\caption{\label{fig:profiles2} Excess kinetic energy and density profiles of the granular bed in a system at packing $\phi=0.75$, forcing amplitude $A_0=1.5\sigma_g$, and orientation $\alpha=\pi/2$ in (a, c), and $\alpha=0$ in (b, d). Positive values of $h$ denote the outer region of the dimer while negative values of $h$ denote the region between inclusions. In hollow points we present the result of the superposition of independent profiles with origin at $h=0$, and $h=-d$. In cyan we plot the relative excess density profile, practically zero but with in (a, c) close to the inclusions.}
\end{figure}

In Fig.~\ref{fig:profiles2} we display the relative excess of density, and kinetic energies for $\phi=0.75$, $A_0=1.5\sigma_g$ relative to the bulk values (\ref{eq:de}).
We compute the profiles for parallel and perpendicular configurations of the dimer and two different separation distances, $d=2\sigma$ and $d=6\sigma_g$.
For positive values of $h$, the behavior of each profile does not depend on the dimer separation $d$.
To confirm this dependence we have computed the profile for a sole inclusion in the system and obtained the same external profiles.

We do not observe a significant deviation of the averaged density along the axis. Moreover, in the region between inclusions, the local density equals to the averaged density outside --except for the density at contact that decays due to the sampling size and the curvature of the disks.
However, both kinetic energies --absolute, and relative-- are not kept constant along $h$.
On the one hand, the excess relative kinetic energy $\delta t_k$ has a positive value at contact and relaxes to zero.
On the other hand, the excess of absolute kinetic energy departs from negative value and relaxes to zero.
For orientations of the dimer perpendicular to the external shaking, $\delta e_k$ jumps to a positive value and follows the decay of $\delta t_k>0$ and relaxes to 0,
whereas for parallel configurations $\delta e_k$ does not follow the relative kinetic energy but slowly increases to zero keeping its negative sign.

Finally, we have analyzed the additivity of one-inclusion profiles for the inner region of the dimer, $h<0$.
From the disturbance profile generated by a single particle $\delta p(h)$, as presented in (\ref{eq:de}), we compute the total disturbance profile in the inner region between independent inclusions located at $h=0$, and $h=-d$  as $\delta p(-h) + \delta p(d-h)$.
We show the additivity of the profiles by comparing the hollow squares, and dots to the solid lines in Fig.~\ref{fig:profiles2}, resulting from the addition of one-inclusion profiles and the computation of the profiles for two-inclusion systems, respectively. 

\section{\label{sec:conclusions} Conclusions}
In summary, we have introduced a mechanical model to capture the effective behavior of inclusion in a horizontally shaken granular bed.
The analysis of the behavior of an inclusion pair has shown they interact with a granular mediated non-central, and long range interactions.
We have considered both a freely moving pair and have computed the forces for a pair at a fixed distance.
These two approaches have provided complementary insight.

We have extracted the two dimensional probability density of the relative vector between inclusions and measured a preference of dimers at contact to align parallel to the external forcing.
Consistently, the force measurements with fixed inclusions have revealed the emergence of an attractive force with a maximum for particles at contact plus the emergence of an aligning torque at short distances.
The detailed computation of the forces between the two inclusions as a function of the angle they form with the driving force indicates that inclusions tend to remove the ordered structure of the granular bed for $\alpha<\pi/2$ and moderate amplitudes of the shaking $A_0>\sigma_g/2$.
When the granular structure around the inclusions becomes weak in the internal region of the pair, the long range attractive force between the pair becomes apparent.
We have characterized the interaction in terms of the parameters of the dimer configuration and shaking of the granular bed with a simple model of interaction at a distance (\ref{eq:independence}).
We have matched the anisotropy in the probability $\mathcal{P}(r_x,r_y)$ to the appearance of a neat torque in for fixed inclusions.
The torque aligns close inclusions $d\lesssim 2\sigma_g$ parallel to the shaking whereas at $d\gtrsim 4\sigma_g$ and large shaking amplitudes and granular densities an emergent torque aligns inclusions perpendicularly to the shaking.
The appearance of this dealigning torque could be the mechanism leading towards the segregation of mixtures of grains and inclusions into bands perpendicular to the shaking direction\cite{Mullin2000,Reis2002,Reis2004,Ciamarra2006b}.
Finally, we have captured the kinetic energy profiles in the granular bed in the immediacies of the inclusion dimer and observed a differentiated behavior of the absolute kinetic energy on the orientation of the dimer.
Additionally we observe that the perturbation of the granular kinetic energies in the internal region of the dimer results from the addition of two independent facing inclusion surfaces

\begin{acknowledgments}
The authors acknowledge I. Zuriguel for  a detailed description of the experimental setup and insightful discussions.
This work has been supported by MINECO and DURSI projects PGC2018-098373-B-100 and 2017 SGR 884, respectively.
JC was supported by Spanish MECD fellowship FPU13/01911.
IP also acknowledges SNF Project No. 200021-175719 for funding.
\end{acknowledgments}

\bibliography{biblio}

\appendix

\section{\label{ap:sdg}Stochastic Dynamics of Grains}
We describe the dynamics of granular particles at positions $\bm r$ and momenta $\bm p$
\begin{equation}
\frac{d}{dt} \bm r_i(t) = \frac{1}{m_i}\bm p_i(t),
\label{eq:ap1}
\end{equation}
where $m$ is the mass of each granular particle. The dynamics are given by the following Langevin set of equations
\begin{equation}
\frac{d}{dt}\bm p_i = \bm F_i -\gamma_i \left(\frac{\bm p_i}{m_i}-\bm v_s\right) + \bm f_i,
\label{eq:langevin_p}
\end{equation}
where $\gamma$ are the dissipation constants, $\bm v_s$ the velocity of the substrate, and $\bm F$, and $\bm f$ the deterministic and random forces, respectively.

To solve equation (\ref{eq:langevin_p}) we separate the terms that depend on the momenta $\bm p$ on LHD separate the deterministic $\bm \Phi=\bm F + \gamma \bm v_s $, and random forces $\bm f_i$ on the RHS
\begin{equation}
\left(\frac{d}{dt} + \frac{\gamma_i}{m_i}\right)\bm p_i = \bm F_i + \gamma_i\bm v_s + \bm f_i = \bm \Phi_i + \bm f_i.
\label{eq:langevin_p2}
\end{equation}
The solution to equation (\ref{eq:langevin_p2}) is computed as $\bm p = \bm p^{(h)}+\bm p^{(p)}$ where homogeneous solution is,
\begin{equation}
\bm p^{(h)}_i\left(t+t_0\right) = \bm p_i(t_0) \exp\left[\frac{\gamma_i}{m_i}\left(t-t_0\right)\right],
\end{equation}
and a particular solution, $\bm p^{(p)}(t+t_0)$, of the form, $\bm \pi(t) \exp \left(-\gamma / m t\right)$, where $\bm \pi(t)$ satisfies the following random process,
\begin{equation}
\frac{d}{dt}\bm \pi_i(t) = \exp\left(\frac{\gamma_i}{m_i}t\right)\left(\bm \Phi_i + \bm f_i\right),
\label{eq:langevin_p3}
\end{equation}
and the solution
\begin{eqnarray}
\bm \pi_i \left(t+\Delta t\right) & -& \bm \pi\left(t\right) = \nonumber\\
  &=&\frac{m_i}{\gamma_i} \left[\exp\left(\frac{\gamma_i}{m_i}\Delta t\right)-1\right]\bm \Phi_i + \\
 &+&\int_0^{\Delta t} d\tau \exp\left[-\frac{\gamma_i}{m_i}\left(\Delta t - \tau \right)\right]\bm f_i(\tau)~.\nonumber
\end{eqnarray}
To compress the result we define a new set of constants. The dissipation of moment is introduced by $\Gamma_i=\exp\left(-\gamma_i \Delta t  / m_i \right)$; the mobility coefficient for the deterministic force $\tilde{\mu}_i = \left(1-\Gamma\right)/\gamma_i$; the effective diffusion coefficient $\tilde{D}_i = \left(\Gamma_i m_i / \gamma_i\right)^{1/2}\left(1-\Gamma_i^2\right)^{1/2}$. Now the evolution of $\bm p$ is simplified to,
\begin{equation}
\bm p_i(t+\Delta t) = \Gamma_i \bm p_i(t) + \tilde{\mu} m_i \bm \Phi_i + \tilde{D}_i \bm \zeta_i,
\label{eq:evolution_p}
\end{equation}
with a Gaussian random noise with zero mean and variance $\left\langle \zeta_{i,\alpha}(t)\zeta_{j,\beta}(t't)\right\rangle = 2 \Lambda_{i,\alpha} \delta_{ij}\delta_{\alpha\beta} \delta (t-t')$. The evolution of the grains is simplified to
\begin{equation}
\bm r_i(t+\Delta t) = \bm r_i(t) + \frac{3\bm p_i(t) + \bm p_i(t+\Delta t) }{4m_i}\Delta t ~.
\label{eq:granular-integration}
\end{equation}
where we first integrate equation (\ref{eq:ap1}) with $\bm p(t)$ to $t+\Delta t/2$, and second we integrate the positions $\bm r(t+\Delta t/2)$ to $t+\Delta t$ with momenta given by $(\bm p(t)+\bm p(t+\Delta t)/2$ from eq. (\ref{eq:evolution_p}).
\section{\label{ap:densitats}Granular measures}
In order to characterize the granular bed we introduce the measure of kinetic energies.
For a system under shaking we can construct either an absolute, and a relative kinetic energy,
\begin{equation}
e_k =  \frac{1}{N}\sum_i \frac{m_i}{2} \left| \bm v_i\right|^2  , ~
t_k  = \frac{1}{N}\sum_i \frac{m_i}{2} \left|\bm v_i-\bm V_{cm}\right|^2,
\label{eq:kinetic}
\end{equation}
with $\bm V_{cm}$ the centre of mass velocity.
The usual magnitude~\cite{Song2005} defined in granular systems is $t_k$ but the introduction of external bodies gives a certain relevance to $e_k$.
Here we discuss the advantages of using each of them to scale the interaction forces obtained in Section~\ref{sec:fixed}.

\begin{figure}[h!]
	\centering
	\includegraphics[scale=1]{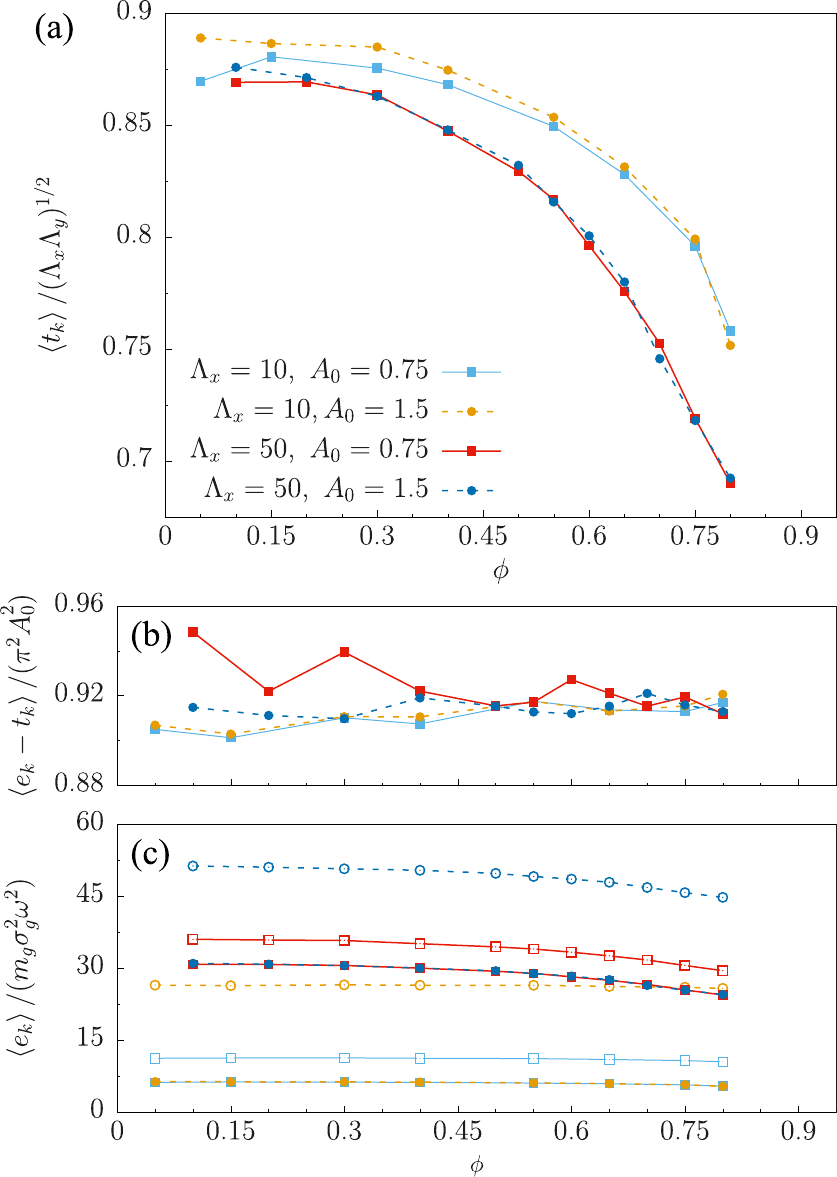}
	\caption{\label{fig:densitats} Measures of kinetic energies as for increasing packing density of grains. In solid lines we plot the measures. In (a) the relative kinetic energy normalized by the random force. In (b) the difference in kinetic energies expected to be $\left\langle e_k-t_k\right\rangle \approx  \pi^2A^2_0$. In (c) the absolute kinetic energy of the grains compared to the acceleration unit given by $m_g\sigma_g^2\omega^2$}
\end{figure}

The relative kinetic energy $t_k$ does not depend on the shaking amplitude $A_0$ and it is best suited to compare forces in systems at different shaking amplitudes and granular properties such as MSD.

The absolute kinetic energy $e_k$ includes the flux of energy that the shaking introduces to the grains, and it is relevant for the interactions with external bodies. Even though both energies are suitable to define an energy scale we choose the averaged $t_k$ for the sake of commensurability.

In Fig.~\ref{fig:densitats} we present the dependence of the averaged energies measured at different densities, forcing amplitudes, and $\Lambda_x$, for a system with fixed granular parameters: $\Lambda_x/\Lambda_y=2$,~$k=2.5\ 10^4$,~ $\gamma_s=20$, and $\gamma=33$ in system units. 

We observe a decay of the kinetic energy as the density of grains increases.
Dissipation in particle-particle collision events in granular systems introduces an energy loss and it is responsible for the energy dissipation of grains as density increases.
The difference in kinetic energies is solely due to the external shaking and it is confirmed in Fig.~\ref{fig:densitats}(b) where the the kinetic energy difference remains approximately constant in the considered packing fraction range. We determine a monolayer to be dense when the energy loss is relevant, in this case for $\phi>0.6$. Since we are interested in a dense system we determine our lower bound for the packing range to be $\phi=0.6$

The upper bound for the packing fraction must correspond to a system within the same mobility regime.
For this purpose we measure the jump distribution of grains, $P(\Delta)$ as the probability density distribution of the displacement $\Delta$ of grains after a time unit.
\begin{figure}[h]
	\centering
	\includegraphics[scale=1]{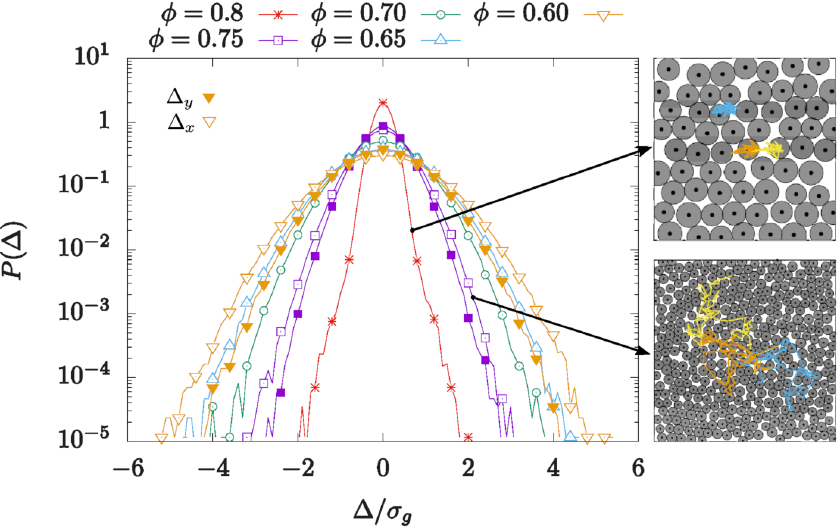}
	\caption{\label{fig:dist_jumps} Displacement statistics of granular particles after between shaking cycles. Plots of the system with three representative trajectories.}
\end{figure}

Fig.~\ref{fig:dist_jumps} shows a Gaussian jump distributions for $\phi\leq 0.75$ at $\Lambda_x=50$. For this reason we set the packing upped limit for exploration in this paper at $\phi=0.75$.

At larger densities, for example, $\phi=0.8$ the jump distribution presents a pronounced peak at $\vert\Delta\vert \approx 0$ that indicates the caging effect of the neighboring grains and it is no longer Gaussian. Mean Squared Displacements, not shown here, also confirm the caging of grains, and thus the system is not in a dense fluidized regime.

\end{document}